\documentclass[conference]{IEEEtran}
\IEEEoverridecommandlockouts
\usepackage{cite}
\usepackage{amsmath,amssymb,amsfonts}
\usepackage{graphicx}
\usepackage{textcomp}
\usepackage{xcolor}
\def\BibTeX{{\rm B\kern-.05em{\sc i\kern-.025em b}\kern-.08em
    T\kern-.1667em\lower.7ex\hbox{E}\kern-.125emX}}
\usepackage{algorithmicx}
\begin{document}

\title{A MARL Based Multi-Target Tracking Algorithm Under Jamming Against Radar \thanks{This work was supported by Tsinghua University Initiative Scientific Research Program under Grants 20234180184.}}
\author{
\IEEEauthorblockN{Ziang Wang \qquad Lei Wang \qquad Qi Yi \qquad Yimin Liu}
\IEEEauthorblockA{Department of Electronic Engineering, Tsinghua University, Beijing, China}
}

\maketitle

\begin{abstract}
Unmanned aerial vehicles (UAVs) have played an increasingly important role in military operations and social life. Among all application scenarios, multi-target tracking tasks accomplished by UAV swarms have received extensive attention. However, when UAVs use radar to track targets, the tracking performance can be severely compromised by jammers. To track targets in the presence of jammers, UAVs can use passive radar to position the jammer. This paper proposes a system where a UAV swarm selects the radar's active or passive work mode to track multiple differently located and potentially jammer-carrying targets. After presenting the optimization problem and proving its solving difficulty, we use a multi-agent reinforcement learning algorithm to solve this control problem. We also propose a mechanism based on simulated annealing algorithm to avoid cases where UAV actions violate constraints. Simulation experiments demonstrate the effectiveness of the proposed algorithm. 
\end{abstract}


\section{Introduction}
\label{sec:intro}
Unmanned aerial vehicles (UAVs) are often used in a multi-target tracking (MTT) scenario due to their easy deployment, low costs, and high flexibility \cite{7470933}. This scenario has been widely studied \cite{8337901,8506387,10158783} because of its crucial role in public safety and military applications. When the tracking distance is long, radars are often used in this scenario to sense target state \cite{10356137}\cite{10476610} thanks to its long detection range and all-weather capability. 

In real-world scenarios, especially in military applications, tracking performance of radars is often seriously affected by malicious jammers \cite{10448275}. To the best of the authors' knowledge, no study has considered how UAVs can avoid degradation of tracking performance in the presence of jammers against radar. Although researchers have explored the way to avoid jamming against inter-UAV communications through movement decisions and frequency selection \cite{10319327}, there is a significant difference between the two cases. The difference exists not only because of the double distance attenuation of the radar's signal-to-noise ratio (SNR), but also because the radar can utilize jamming signals to position jammers. Considering the commonness of the scenario, this gap needs to be filled urgently. 

For targets with jammers on them, a common method of tracking is to measure the direction of arrival (DOA) of the jamming signal by different radars through passive reception. However, radar's passive reception and active emission are often contradictory. To solve this problem, we propose an anti-jamming method, where UAV needs to decide on its own movement in addition to selecting radar work mode between active mode (AM) and passive mode (PM).

For UAV-MTT problem,  multi-agent reinforcement learning (MARL) is a commonly used method \cite{10198223,9406813,9963915,9623508,9446301,9839387}. Some studies consider multi-UAV search and tracking of collaborative beacons. The researchers in \cite{10198223} considered how multiple UAVs in a discrete grid-world can simultaneously accomplish modeling of environmental obstacles and tracking of moving beacons using MARL. The authors of \cite{9406813} consider a continuous environment where UAVs track moving beacons, but the UAVs’ action is simply choosing one of several fixed directions to move a fixed distance. More researches focus on the non-cooperative tracking problem, where it may have malicious jammers. Researchers in \cite{9963915} consider UAV detection of nearly fixed targets using MARL. When it comes to moving targets,  In \cite{9623508}, the authors consider how multiple UAVs trade-off for both saving flight energy and performing target tracking, and propose an energy-efficient UAV-MTT algorithm. In \cite{9446301}, the authors consider the MTT scenario without obstacles in which each UAV serves as an integrated sensing and communication (ISAC) system. Besides, researchers in \cite{9839387} explored how to allow UAVs to deliver valuable information to neighbors through fewer communications to enable pre-scheduling in MTT. As mentioned earlier, existing studies have not focused on UAV-MTT in the presence of jammers. 

In addition, UAV actions are often subject to a number of constraints determined by the scenario. Despite penalizing constraint violations in reward function, most studies cannot completely prevent constraints from being violated \cite{10446347}. How to prevent UAVs from violating constraints without affecting their performance is also a problem that needs to be solved.


In this paper, we model the UAV-MTT problem in the presence of jammers against radar, in which each radar decide on its own work mode (AM or PM) based on local observation. We formulate the optimization problem and prove the difficulty of solving it directly. Then we implemented a MARL algorithm to solve the problem. We also propose an extra mechanism based on simulated annealing (SA) algorithm to prevent UAV actions from violating constraints while maintaining tracking performance. Finally, we verify the performance of the proposed algorithm by simulation. 

The rest of this paper is organized as follows: Section 2 presents the system model, Section 3 describes the MARL model and the proposed algorithm, Section 4 shows the simulation results and Section 5 shows our main conclusions.

\section{SYSTEM MODEL}
\label{sec:system}

\subsection{Scenario Overview}
\label{ssec:scenario}

We consider a UAV-MTT scenario shown in \textbf{Fig.1}, where $N$ UAVs track $M$ targets in $T$ timesteps. The UAV swarm jointly decide on target assignment, trajectory decision, and radar work mode for each UAV. The location of UAV $i$ at timestep $k$ is noted as $\mathbf{p}_{T_k}^i=(x_{T_k}^i, y_{T_k}^i)^T$, and position of target $j$ is $\mathbf{p}_{T_k}^j=(x_{T_k}^j, y_{T_k}^j)^T$. The set of target indexes with jammer is $\mathcal{J}^\text{Y}_{T_k}$, while $\mathcal{J}^\text{N}_{T_k}$ means the contrary. As is mentioned before, the set of UAVs in AM is $\mathcal{U}^{\text{AM}}_{T_k}$ sized $n_{T_k}^\text{AM}$, and the set of UAVs in PM is $\mathcal{U}^{\text{PM}}_{T_k}$ sized $n_{T_k}^\text{PM}$. Radar work mode is denoted as $u_{T_k}^i \in \{0, 1\}$, with $0$ for PM and $1$ for AM. When the radar chooses PM, multiple radars achieve cooperative positioning by measuring DOA of the same jamming signal. 

\begin{figure}[htb]
\begin{center}
\includegraphics[width=8.6cm]{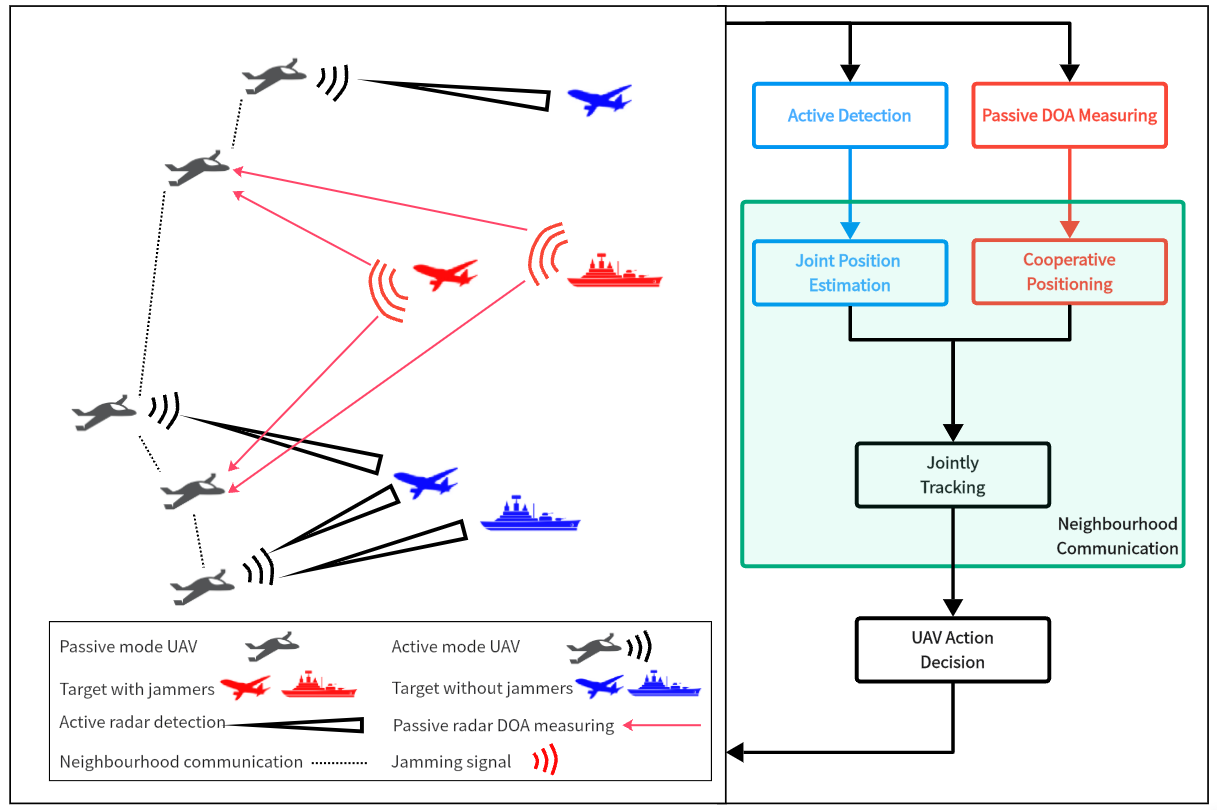}
\caption{Scenario schematic and system workflow}
\end{center}
\end{figure}

\subsection{Performance Indicator}
\label{ssec:obj}

In the scenario described previously, UAV actions controlled by the algorithm are
$\mathbf{a}_{T_k}^i = (\Delta x_{T_k}^i, \Delta y_{T_k}^i, u_{T_k}^i)^T$, where 
\begin{gather}
    \begin{bmatrix} \Delta x_{T_k}^i \\ \Delta y_{T_k}^i \end{bmatrix} = 
    \begin{bmatrix} x_{T_k}^i-x_{T_{k-1}}^i \\ y_{T_{k}}^i-y_{T_{k-1}}^i \end{bmatrix}.
\end{gather}
For ease of presentation, we will subsequently default to this relationship and use $\mathbf{p}_{T_k}^i$ to represent UAV movement.

To represent the performance of target tracking, we calculate the average Cramer Rao's Lower Bound (CRLB) of position estimation of each target. For targets with or without jammers, the CRLB is determined by the UAVs in PM or AM, respectively. 

Specifically, for a target $j$ without jammer, each AM radar $i$ can get an observation as \cite{2017Cooperative}: 
\begin{gather}
    \begin{bmatrix} r_{i, j, k} \\  \theta_{i, j, k} \end{bmatrix} = 
    \begin{bmatrix} \sqrt{(x_{T_k}^i-x_{T_k}^j)^2+(y_{T_k}^i-y_{T_k}^j)^2} 
    \\ \text{arctan}(\frac{y_{T_k}^j-y_{T_k}^i}{x_{T_k}^j-x_{T_k}^i}) \end{bmatrix} + \mathbf{n_{i,j,k}},
\end{gather}
where $\mathbf{n_{i,j,k}} \sim 
N(\mathbf{0}, \text{diag}(\frac{r_{i, j, k}^4}{f_{i, j, k, 1}}, \frac{r_{i, j, k}^4}{f_{i, j, k, 2}}))
$
is an Additional White Gaussian Noise (AWGN), $f_{i, j, k, 1}$ and $f_{i, j, k, 2}$ are factors related to radar transmit power, frequency, bandwidth, gain, and target Radar Cross-Section (RCS), etc., and their specific expressions can be found in \cite{2017Cooperative}. For a given radar and target, considering the average effect over a period of time, these factors can be approximated as constants without loss of generality. The estimation CRLB of 
$(x_{T_k}^j, y_{T_k}^j)^T$ at the $k$th timestep can be expressed as:
\begin{gather}
    \text{CRLB}^{-1}_1(\mathbf{p_{T_k}^j}) = \sum_{\substack{i=1, \\i \in \mathcal{U}^{\text{AM}}_{T_k}}}^N \textbf{H}^T_{1,i,j,k} \mathbf{\Sigma}^{-1}_{i,j,k} \textbf{H}_{1,i,j,k},
\end{gather}
where $\textbf{H}_{1,i,j,k} = \begin{bmatrix} \frac{x_{T_k}^j-x_{T_k}^i}{r_{i, j, k}}&\frac{y_{T_k}^j-y_{T_k}^i}{r_{i, j, k}}\\
\frac{y_{T_k}^i-y_{T_k}^j}{r^2_{i, j, k}}&\frac{x_{T_k}^j-x_{T_k}^i}{r^2_{i, j, k}}\end{bmatrix}$, and $\mathbf{\Sigma}_{i,j,k}=\text{diag}(\frac{r_{i, j, k}^4}{f_{i, j, k, 1}}, \frac{r_{i, j, k}^4}{f_{i, j, k, 2}})$ is the covariance matrix of $\mathbf{n_{i,j,k}}$. 

For a target with jammer, PM radars get DOAs of the jamming signals with an AWGN to estimate the position of the jammer. The position estimation CRLB is: 
\begin{gather}
    \text{CRLB}^{-1}_2(\mathbf{p_{T_k}^j}) = \textbf{H}^T_{2,j,k} \mathbf{\Sigma}^{-1}_{j,k} \textbf{H}_{2,j,k},
\end{gather}
where $\textbf{H}_{2,j,k}$ has $n^\text{PM}_{T_k}$ rows and 2 columns. The $i$ th row of it is $\begin{bmatrix} \frac{y_{T_k}^{n(i)}-y_{T_k}^j}{r^2_{n(i), j, k}}&\frac{x_{T_k}^j-x_{T_k}^{n(i)}}{r^2_{n(i), j, k}} \end{bmatrix}$, where $n(i)$ is the $i$ th UAV index in $\mathcal{U}^{\text{PM}}_{T_k}$. Matrix $\mathbf{\Sigma}_{j,k}$ is a diagonal matrix, and the $i$th diagonal element is $\frac{r_{n(i), j, k}^2}{f_{n(i), j, k}}$ denoting the DOA measurement noise variance of radar $n(i)$ with factor $f_{n(i), j, k}$ similar to the previous two factors.

Finally, we get the multi-target average CRLB to indicate the precision of coordinate estimation at timestep $T_k$ as:
\begin{gather}
    \frac{1}{M}[\sum_{\substack{j=1, \\j \in \mathcal{J}^\text{N}_{T_k}}}^M \text{tr}(\text{CRLB}_1(\mathbf{p_{T_k}^j}))+\sum_{\substack{j=1, \\j \in \mathcal{J}^\text{Y}_{T_k}}}^M \text{tr}(\text{CRLB}_2(\mathbf{p_{T_k}^j}))].
\end{gather}
 We define it as $\text{LB}_{T_k}$ to show the tracking variance. Due to the different UAV-target distances, the order of magnitude of $\text{LB}_{T_k}$ may be different for different k. Therefore, we calculate the geometric mean of the whole time series to uniformly reflect the overall tracking effect, which means $\text{TE} = -\sum_{k=1}^Tlg(\text{LB}_{T_k})$. The negative sign is used to maintain a positive correlation between the value of TE and the real tracking effect. 

\subsection{Constraints}
\label{ssec:constraints}
 For each two UAV $i, i' \in \{1,2,...,N\}$ and target $j$,  There are constraints on their positions that must be observed. In this paper, we consider the following three constraints based on the requirements of realistic scenarios:
\begin{gather}
||\textbf{p}^i_{T_k} - \textbf{p}^i_{T_{k-1}}|| \leq d_0, \\
||\textbf{p}^i_{T_k} - \textbf{p}^{i'}_{T_k}|| \geq d_1, \\
||\textbf{p}^i_{T_k} - \textbf{p}^j_{T_k}|| \geq d_2,
\end{gather}
where $|| \cdot ||$ is the 2-norm, $d_0, d_1$ and $d_2$ are known arguments. Constraint (6) indicates that the UAV has limited mobility. 
Constraint (7) limits the minimum distance between different UAVs to prevent collisions. And constraint (8) limits the minimum distance between a UAV and a target to prevent attack from hostile targets. 

\section{Algorithm}
\label{sec:algo}

In \textbf{Sec \ref{sec:system}}, we summarize a problem that can be optimized from the real scenario. However, feasible set determined by (7) and (8) is not convex. Therefore, the optimization cannot be solved directly. For this reason, we use a heuristic algorithm, MARL, to solve the optimization problem. 

\subsection{Dec-POMDP}
\label{ssec:pomdp}
In a MARL algorithm, we consider the environment as a decentralized partially observable Markov decision process (Dec-POMDP)\cite{2016} defined by a tuple $<\mathcal{S, A}, O, R, P, n, \gamma>$. Set $\mathcal{S}$ is the state space containing UAV position $\mathbf{p_{T_k}^i}$ and target position $\mathbf{p_{T_k}^j}$ for all $i, j$, jammer situation set $\mathcal{J}^\text{Y}_{T_k}$, $\mathcal{J}^\text{N}_{T_k}$ and lastest work mode set $\mathcal{U}^\text{AM}_{T_{k-1}}$, $\mathcal{U}^\text{PM}_{T_{k-1}}$. Set $\mathcal{A}$ is the action space i.e. action vectors. 
Function $O$ is observation function of each agent and $P$ is the transition posibility of state from $T_{k-1}$ to $T_k$ with actions given. In our problem, each agent can observe positions of other UAVs and the predicted positions of targets at next timestep. Both types of position are relative position to the agent. Note that target motion patterns are not modeled by our algorithm , and target next positions are predicted by a separate tracking module. Function $R$ is the reward function. In our algorithm, for each agent, the reward $R_{i, T_k}$ can be split into shared reward $R^\text{s}_{ T_k}$ and distinct reward $R^\text{d}_{i, T_k}$. $R^\text{s}_{T_k}$ is the same for different agents. It negatively correlates with $\text{LB}_{T_k}$. $R^\text{d}_{i, T_k}$ is the difference between shared reward with all agents and that without agent $i$. It is designed to prevent "lazy" agents. In addition, penalty $M_{i, T_k}$ for violating constraint (7) and (8) is subtracted from reward. This value is 0 when the above constraints are not violated. The form of $R_{i, T_k}$ is as follows: 
\begin{gather}
R_{i, T_k} = R^\text{s}_{T_k} + \alpha R^\text{d}_{i, T_k} - M_{i, T_k},
\end{gather}
where $\alpha$ is a constant.
Finally, $n$ is the number of agents, i.e. $N$ in our problem and $\gamma$ is the discount factor. 

\subsection{Constraint MARL Algorithm}
\label{ssec:learning}
We use MAPPO algorithm \cite{MAPPO} to solve the problem. However, a drawback of the original MARL algorithm is that even large penalties does not guarantee that the constraints are obeyed \cite{10446347}. Researchers in \cite{10446347} propose a mechanism to deal with the problem. However, the mechanism is primarily used to identify violations of constraints and is not concerned with the way to adjust the actions without reducing the tracking effect. To deal with this gap, we propose a mechanism based on SA algorithm to tune the actions.

Specifically, in our problem, constraint (6) is naturally obeyed. As for constraint (7), the collision distance of UAVs ($d_1$) is always close, so the problem is easy to dealt with by using obstacle avoidance algorithms. Therefore, we only focus on the treatment of violations of constraint (8). 

After acquiring actions $\{\mathbf{a}_{T_k}^i|i \in \{1,2,...,N\}\}$, we compute $\hat{\textbf{p}}^i_{T_k} = \textbf{p}^i_{T_{k-1}} + \Delta \textbf{p}^i_{T_k}$. Given the predicted position of target $j$ as  $\hat{\textbf{p}}_{T_k}^j$, we can compute $||\hat{\textbf{p}}^i_{T_k} - \hat{\textbf{p}}^j_{T_k}||$. If the value is less than $d_2 + 3\sigma_\text{pred}$, where $\sigma_\text{pred}$ is the standard deviation of prediction, we invoke a SA module to tune the action. In the SA module, the value function to be minimized is: 
\begin{gather}
R_{i, T_k} = -( \hat{R}^\text{s}_{i, T_k} - \hat{M}'_{i, T_k} - L \cdot I(||(\Delta x_{T_k}^i, \Delta y_{T_k}^i)|| < d_0)),
\end{gather}
where $\hat{R}^s_{i, T_k}$ is the shared reward calculated using the predicted position of this UAV $\hat{\textbf{p}}_{i, T_k}$ got by executing the optimized action, the last state of other UAVs and the predicted state of targets. $L$ is a sufficiently large value to ensure that actions satisfy constraint (6), $I$ is an indicator function, and $\hat{M}'_{i, T_k}$ is a positive constant when: 
\begin{gather}
{\exists} j \in \{1,...,M\}, s.t. ||\hat{\textbf{p}}_{i, T_k} - \hat{\textbf{p}}_{j, T_k}|| < d_2 + 3\sigma_\text{pred}.
\end{gather}
If this condition is not satisfied, $\hat{M}'_{i, T_k}$ is 0. 

The complete algorithm flow is shown in \textbf{Fig.2}, where the SA solving process is highlighted with a red border. 

\begin{figure}[htb]
\begin{center}
\includegraphics[width=8.6cm]{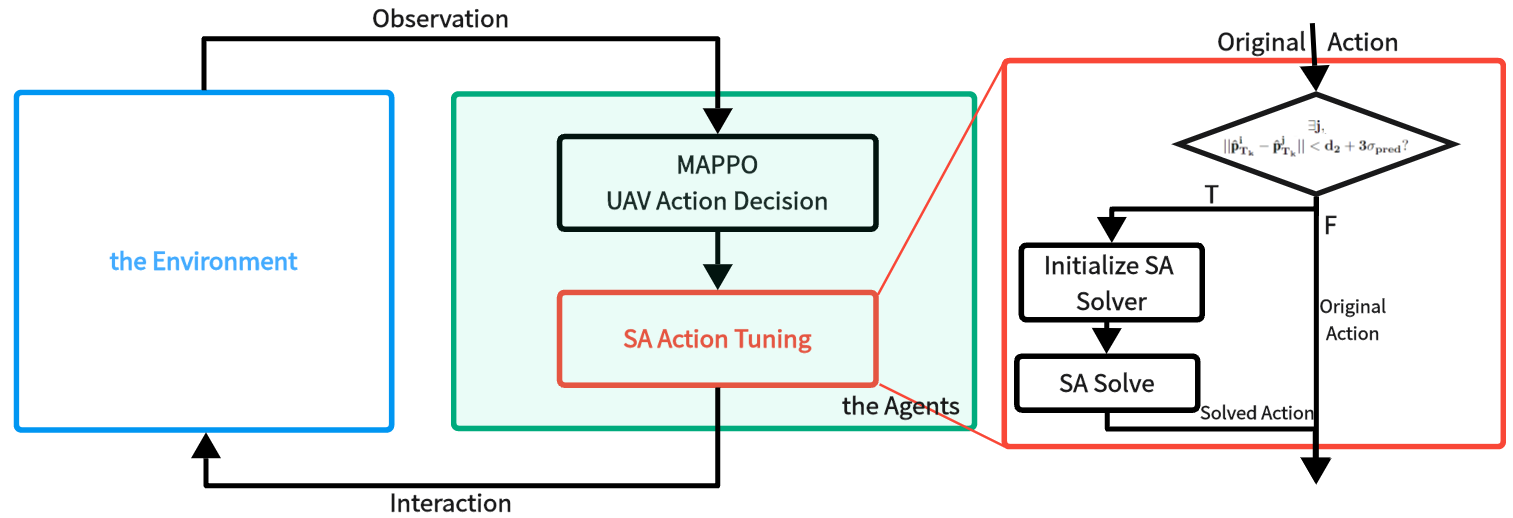}
\caption{Algorithm flow}
\end{center}
\end{figure}

\section{Results}
\label{sec:results}
In the numerical simulation, we consider $N = 6$ UAVs track $M = 3$ targets. The probability of a target carrying a jammer is set to be 0.5. Initial positions of all UAVs are fixed around the coordinate origin without loss of generality. Initial positions of targets have a fixed distribution in polar coordinates. The distance from target initial positions to the coordinate origin is sampled from a uniform distribution $U(r_\text{min}, r_\text{max})$. In a realistic battlefield environment, targets often arrive from a certain direction. Therefore, the initial angle of the target is set to $U(\theta_\text{init}+\theta_\text{min}, \theta_\text{init}+\theta_\text{max})$, where $\theta_\text{init}$ is randomly sampled from $U(0, 2\pi)$ in radian to maintain the rotational symmetry of the algorithm's performance. 
In our simulation, targets take a uniform motion with driven noise. Target initial velocity magnitude is sampled from $U(r_\text{v,min}, r_\text{v,max})$, and velocity angle is sampled from $U(\theta_\text{init}+\theta_\text{v,min}, \theta_\text{init}+\theta_\text{v,max})$. 

Every episode contains 300 steps. As for the SA model, we have $T_\text{max}=100$ and $T_\text{min}=20$. In this temperature range, the transition from constraint-satisfying actions to constraint-violating actions is virtually impossible, but for transitions between actions that satisfy constraints, the algorithm encourages exploration rather than having to find an optimum. This is intended to counteract modeling mismatches in tracking effect and biases caused by using state of last timestep. The iteration rounds in the tuning process is 20. 

We take MAPPO, MATD3 and MADDPG as baselines. The hyperparameters of all algorithms have been tuned to a suitable value through experiments. Actors and critics of all the algorithms are multi-layer perceptrons (MLP) with 5 hidden layers and 256 neurons per layer. The learning rate of the proposed algorithm and MAPPO is $5\times 10^{-5}$ and that of MATD3 and MADDPG is $1\times 10^{-8}$. Results for different algorithms are shown in \textbf{Fig.3}. \textbf{Fig.3(a)} shows the average rewards of each episode, while \textbf{Fig.3(b)} shows TE over each episode to show the real tracking performance. Here, all the episode rewards are averaged over 10 random seeds and 10 episodes for each seed, and smoothed over a window of 50. Our code and other parameters are available at \textbf{https://github.com/s1s3r4/mUAV-MTT-MAPPO}.

\begin{figure}[htb]
\begin{minipage}[b]{.48\linewidth}
  \centering
  \centerline{\includegraphics[width=4.2cm]{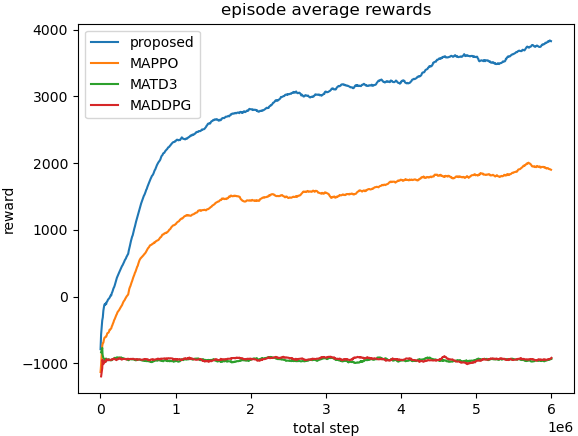}}
  \centerline{(a) Episode average rewards}\medskip
\end{minipage}
\hfill
\begin{minipage}[b]{0.48\linewidth}
  \centering
  \centerline{\includegraphics[width=4.2cm]{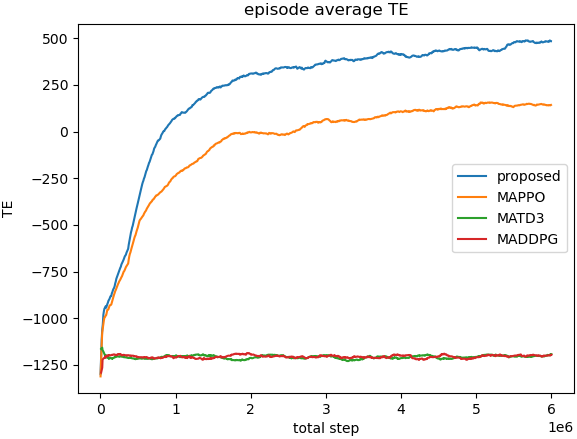}}
  \centerline{(b) Episode average TE}\medskip
\end{minipage}
\caption{Performance of different algorithms}
\end{figure}


In the result, MATD3 and MADDPG, while converging in a short time, fail to learn good decisions due to poor exploration and mismatch between off-policy algorithms and large number of agents. Besides, effect of MAPPO is mainly limited by violation of constraints. Our proposed algorithm is effective in both avoiding violation of constraints and leading to efficient learning, and thus has the best performance. 

Besides, performance of an algorithm can be roughly evaluated by examples. In \textbf{Fig.4}, we give a running example of our algorithm. Here, 3 forks are targets and 6 points are UAVs. Here red forks are targets with active jammers and blue forks are targets without jammers, while red points stand for passive radar UAVs and blue points stand for active radar UAVs. The same color means matched work mode.

\begin{figure}[htb]
\begin{center}
\includegraphics[width=8.6cm]{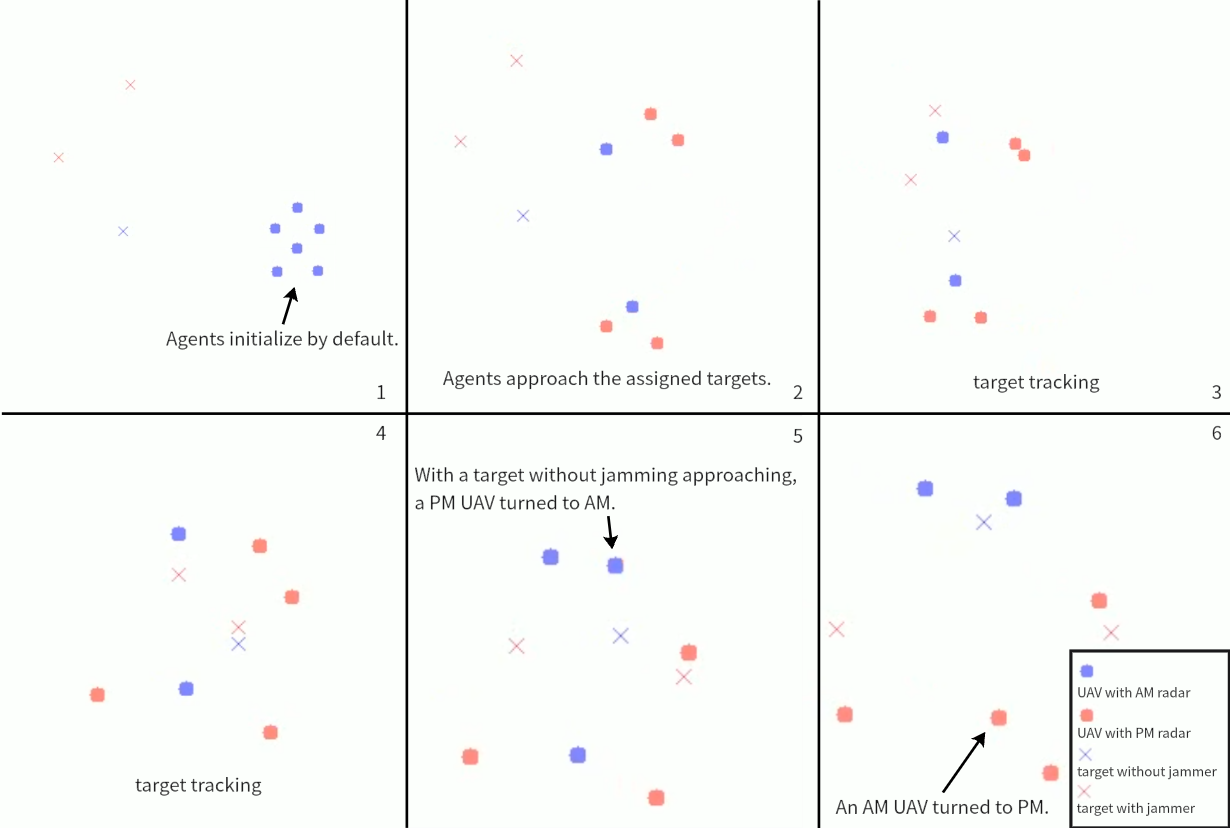}
\caption{A running example of our algorithm}
\end{center}
\end{figure}

The video of this running example can also be found at \textbf{https://github.com/s1s3r4/mUAV-MTT-MAPPO}. The key elements of each subfigure are highlighted in the figure. From \textbf{Fig.4}, it can be seen that the agents' decision is reasonable. 


\section{Conclusions}
\label{sec:conclusions}
In this paper, we consider a scenario where multiple UAVs use radars to track multiple moving targets. Some targets carry jammers against radars. UAVs try to avoid degradation of tracking performance in the presence of malicious jammers and track all targets. We propose a MARL algorithm to control UAVs and a mechanism based on SA to avoid action from violating constraints. Simulations illustrate that the proposed algorithm can effectively improve the performance.

\bibliographystyle{IEEEbib}
\bibliography{refs}

\begin{thebibliography}{10}

\bibitem{7470933}
Yong Zeng, Rui Zhang, and Teng~Joon Lim,
\newblock ``Wireless communications with unmanned aerial vehicles:
  opportunities and challenges,''
\newblock {\em IEEE Communications Magazine}, vol. 54, no. 5, pp. 36--42, 2016.

\bibitem{8337901}
Jingjing Gu, Tao Su, Qiuhong Wang, Xiaojiang Du, and Mohsen Guizani,
\newblock ``Multiple moving targets surveillance based on a cooperative network
  for multi-uav,''
\newblock {\em IEEE Communications Magazine}, vol. 56, no. 4, pp. 82--89, 2018.

\bibitem{8506387}
Arun Das, Shahrzad Shirazipourazad, David Hay, and Arunabha Sen,
\newblock ``Tracking of multiple targets using optimal number of uavs,''
\newblock {\em IEEE Transactions on Aerospace and Electronic Systems}, vol. 55,
  no. 4, pp. 1769--1784, 2019.

\bibitem{10158783}
Wei Xia, Zhuoyang Zhou, Wanyue Jiang, and Yuhan Zhang,
\newblock ``Dynamic uav swarm confrontation: An imitation based on mobile
  adaptive networks,''
\newblock {\em IEEE Transactions on Aerospace and Electronic Systems}, vol. 59,
  no. 5, pp. 7183--7202, 2023.

\bibitem{10356137}
Philipp Stockel, Patrick Wallrath, Reinhold Herschel, and Nils Pohl,
\newblock ``Detection and monitoring of people in collapsed buildings using a
  rotating radar on a uav,''
\newblock {\em IEEE Transactions on Radar Systems}, vol. 2, pp. 13--23, 2024.

\bibitem{10476610}
Yifan Jiang, Qingqing Wu, Wen Chen, and Kaitao Meng,
\newblock ``Uav-enabled integrated sensing and communication: Tracking design
  and optimization,''
\newblock {\em IEEE Communications Letters}, vol. 28, no. 5, pp. 1024--1028,
  2024.

\bibitem{10448275}
Yijia Zhang, Deepak Mishra, Hassan~Habibi Gharakheili, and Derrick Wing
  Kwan~Ng,
\newblock ``Uav operation time minimization for wireless-powered data
  collection,''
\newblock in {\em ICASSP 2024 - 2024 IEEE International Conference on
  Acoustics, Speech and Signal Processing (ICASSP)}, 2024, pp. 46--50.

\bibitem{10319327}
Lanhua Xiang, Fengyu Wang, Wenjun Xu, Tiankui Zhang, Miao Pan, and Zhu Han,
\newblock ``Dynamic uav swarm collaboration for multi-targets tracking under
  malicious jamming: Joint power, path and target association optimization,''
\newblock {\em IEEE Transactions on Vehicular Technology}, vol. 73, no. 4, pp.
  5410--5425, 2024.

\bibitem{10198223}
Anna Guerra, Francesco Guidi, Davide Dardari, and Petar~M. Djuric,
\newblock ``Reinforcement learning for joint detection \& mapping using dynamic
  uav networks,''
\newblock {\em IEEE Transactions on Aerospace and Electronic Systems}, pp.
  1--16, 2023.

\bibitem{9406813}
Jiseon Moon, Savvas Papaioannou, Christos Laoudias, Panayiotis Kolios, and
  Sunwoo Kim,
\newblock ``Deep reinforcement learning multi-uav trajectory control for target
  tracking,''
\newblock {\em IEEE Internet of Things Journal}, vol. 8, no. 20, pp.
  15441--15455, 2021.

\bibitem{9963915}
Tao Zhang, Kun Zhu, Shaoqiu Zheng, Dusit Niyato, and Nguyen~Cong Luong,
\newblock ``Trajectory design and power control for joint radar and
  communication enabled multi-uav cooperative detection systems,''
\newblock {\em IEEE Transactions on Communications}, vol. 71, no. 1, pp.
  158--172, 2023.

\bibitem{9623508}
Zhaoyue Xia, Jun Du, Jingjing Wang, Chunxiao Jiang, Yong Ren, Gang Li, and Zhu
  Han,
\newblock ``Multi-agent reinforcement learning aided intelligent uav swarm for
  target tracking,''
\newblock {\em IEEE Transactions on Vehicular Technology}, vol. 71, no. 1, pp.
  931--945, 2022.

\bibitem{9446301}
Longyu Zhou, Supeng Leng, Qiang Liu, and Qing Wang,
\newblock ``Intelligent uav swarm cooperation for multiple targets tracking,''
\newblock {\em IEEE Internet of Things Journal}, vol. 9, no. 1, pp. 743--754,
  2022.

\bibitem{9839387}
Longyu Zhou, Supeng Leng, Qing Wang, and Qiang Liu,
\newblock ``Integrated sensing and communication in uav swarms for cooperative
  multiple targets tracking,''
\newblock {\em IEEE Transactions on Mobile Computing}, vol. 22, no. 11, pp.
  6526--6542, 2023.

\bibitem{10446347}
Leonardo Spampinato, Enrico Testi, Chiara Buratti, and Riccardo Marini,
\newblock ``Madrl-based uavs trajectory design with anti-collision mechanism in
  vehicular networks,''
\newblock in {\em ICASSP 2024 - 2024 IEEE International Conference on
  Acoustics, Speech and Signal Processing (ICASSP)}, 2024, pp. 12976--12980.

\bibitem{2017Cooperative}
Junkun Yan, Wenqiang Pu, Hongwei Liu, Shenghua Zhou, and Zheng Bao,
\newblock ``Cooperative target assignment and dwell allocation for multiple
  target tracking in phased array radar network,''
\newblock {\em Signal Processing}, vol. 141, no. dec., pp. 74--83, 2017.

\bibitem{2016}
Frans~A. Oliehoek and Christopher Amato,
\newblock ``[springerbriefs in intelligent systems] a concise introduction to
  decentralized pomdps || multiagent systems under uncertainty,''
\newblock vol. 10.1007/978-3-319-28929-8, no. Chapter 1, pp. 1--9, 2016.

\bibitem{MAPPO}
Chao Yu, Akash Velu, Eugene Vinitsky, Yu~Wang, and Yi~Wu,
\newblock ``The surprising effectiveness of mappo in cooperative, multi-agent
  games,''
\newblock 2021.

\end{thebibliography}

\end{document}